\documentclass[twocolumn,aps,pra,floatfix]{revtex4-1}

\usepackage{graphicx}

\begin{document}

\title{Nature of 3D Bose Gases near Resonance}
\author{Dmitry Borzov$^{1}$, Mohammad S. Mashayekhi$^{1}$, Shizhong Zhang$^{2}$, Jun-Liang Song$^{3}$ and Fei Zhou$^{1,4}$}
\affiliation{Department of Physics and Astronomy,
The University of British Columbia, Vancouver, B. C., Canada V6T1Z1$^{1}$ \\
Department of Physics, The Ohio-State University, Columbus,Ohio 43210, USA$^{2}$\\
Institute for Quantum Optics and Quantum Information of the Austrian Academy of Sciences, 6020 Innsbruck, Austria$^{3}$\\
Institute for Advanced Studies,Tsinghua University, Beijing 100084, China$^{4}$}
\date{{\small \today}}


\begin{abstract}
In this paper, we explore the nature of three-dimensional Bose gases at large positive scattering lengths via
resummation of dominating processes involving a minimum number of virtual atoms.
We focus on the energetics of the nearly fermionized Bose gases beyond the usual dilute limit.
We also find that an onset instability sets in at a critical scattering length, beyond which the near-resonance Bose gases become strongly coupled to molecules
and lose the metastability. Near the point of instability, the chemical potential reaches a maximum, and the effect of the three-body forces can be estimated to be around a few percent.
\end{abstract}
\maketitle

\section{Introduction}

Recently, impressive experimental attempts have been made
to explore the properties of Bose gases near Feshbach resonance~
\cite{Navon11,Papp08,Pollack09}. In these experiments,
it has been suggested that when approaching resonance from the side
of small positive scattering lengths in the upper branch,
Bose atoms appear to be thermalized within a reasonably short time,
well before the recombination processes set in,
and so form a quasistatic condensate.
Furthermore, the life time due to the recombination processes
is much longer than the many-body time scale set by
the degeneracy temperature.
This property of Bose gases near resonance
and the recent measurement of the chemical potentials for
a long-lived condensate by Navon {\it et al}.~\cite{Navon11} motivate us to make further theoretical investigations on
the fundamental properties of Bose gases at large scattering lengths.

The theory of dilute Bose gases has a long history, starting with the Bogoliubov theory of weakly interacting Bose gases~\cite{Bogoliubov47}.
A properly regularized
theory of dilute gases of bosons with contact interactions
was first put forward by Lee, Huang, and Yang~\cite{Lee57} and later by Beliaev~\cite{Beliaev58,Nozieres90}, who
developed a field-theoretical approach.
Higher-order corrections
were further examined in later years~\cite{Wu59,BHM02}. Since these
results were obtained by applying an expansion in terms of the small parameter $\sqrt{na^3}$ (here $n$ is the density and
$a$ is the scattering length),
it is not surprising that, formally speaking, each of the terms appearing
in the dilute-gas theory diverges when the scattering lengths are extrapolated to
infinity. As far as we know, resummation of these
contributions, even in an approximate way,
has been lacking~\cite{twobody}. This aspect,
to a large extent, is the main reason
why a qualitative understanding of Bose gases near resonance
has been missing for so long.

There have been a few theoretical efforts to understand the Bose gases at large positive scattering lengths.
The numerical efforts have been focused on the energy minimum in truncated Hilbert spaces, which have been argued to be relevant to
Bose gases studied in experiments~\cite{Cowell02,Song09,Diederix11}. These efforts are consistent in pointing out that the Bose gases are nearly fermionized near resonance.
However, there are two important unanswered questions in the previous studies. One is
whether the energy minimum found in a restricted subspace is indeed metastable in the
whole Hilbert space. The other equally important issue is what the role of three-body Efimov physics in the
Bose gases near resonance is.

Below we outline a nonperturbative approach to the long-lived
condensates near resonance.
We have applied this approach to explore the nature of Bose gases
near resonance and to address the above issues.
One concept emerging from this study is that a quantum gas (either fermionic or bosonic) at
a positive scattering length does not always appear to be
equivalent to a gas of effectively repulsive atoms;
 this idea, which we believe has been overlooked in many recent studies,
plays a critical role in our analysis of Bose gases near resonance.

Our main conclusions are fourfold:
~(a) energetically, the Bose gases close to unitarity are nearly {\it fermionized},
i.e., the chemical potentials of the Bose gases
approach that of the Fermi energy of a Fermi gas with equal mass and density;
~(b) an onset instability sets in at a positive critical scattering length, beyond which the Bose gases appear to lose the metastability
as a consequence of the sign change of effective interactions at large scattering lengths;
~(c) because of a strong coupling with molecules near resonance,
the chemical potential reaches a maximum in the vicinity of the instability point;
~(d) at the point of instability, we estimate, via summation of loop diagrams, the effect of three-body forces to be around a few percent.

Feature (a) is consistent with previous numerical calculations~\cite{Cowell02,Song09,Diederix11};
both (b) and (c) are surprising features, not anticipated in the previous numerical calculations
or in the standard dilute-gas theory~\cite{Lee57,Nozieres90}.
Our attempt here is mainly intended to reach an in-depth understanding of the energetics, metastability of
Bose gases beyond the usual dilute limit as well as the contributions of three-body effects.
The approach also reproduces quantitative features of the dilute-gas theory.
In Sec. II and Appendixes A-C, we outline our main calculations and arguments.
In Sec. III, we present the conclusion of our studies.

\section{Chemical potential, Metastability and Efimov effects}

The Hamiltonian we apply to study this problem is

\begin{eqnarray}
H&=&\sum_{\bf k} (\epsilon_{\bf k} -\mu) b_{\bf k}^\dagger b_{\bf  k}
+ 2 U_0 n_0 \sum_{\bf k} b^\dagger_{\bf  k} b_{\bf  k}
\nonumber \\
&+&\frac{1}{2} U_0 n_0\sum_{\bf k}
b^\dagger_{\bf  k}
b^\dagger_{-\bf  k}+
\frac{1}{2}U_0 n_0 \sum_{\bf k} b_{\bf  k}b_{-\bf  k}
\nonumber \\
&+&\frac{U_0}{\sqrt{\Omega}}\sqrt{n_0}
\sum_{{\bf k'},{\bf q}} b^\dagger_{\bf q} b_{\bf  k'+\frac{\bf q}{2}}
b_{-\bf  k'+\frac{\bf q}{2}}+{\rm H.c.}
\nonumber \\
&+&\frac{U_0}{2\Omega} \sum_{{\bf k}, {\bf k'},{\bf q}} b^\dagger_{\bf k+\frac{\bf q}{2}} b^\dagger_{-\bf  k+\frac{\bf q}{2}} b_{\bf  k'+\frac{\bf q}{2}}
b_{-\bf  k'+\frac{\bf q}{2}}+{\rm H.c.}
\end{eqnarray}
Here $\epsilon_{\bf k}={|\bf k|}^2/2m$, and
the sum is over nonzero momentum states.
$U_0$ is the strength of the contact interaction related to the scattering length
$a$ via $U_0^{-1}=m (4\pi a)^{-1} -\Omega^{-1} \sum_{\bf k} (2\epsilon_{\bf k})^{-1}$, and $\Omega$ is the volume.
$n_0$ is the number density of the condensed atoms and $\mu$ is the chemical potential, both of which are functions of $a$ and are to be determined self-consistently.
The chemical potential $\mu$ can be expressed in terms of $E(n_0,\mu)$,
the energy density for the Hamiltonian in Eq.~(1), with $n_0$ fixed~\cite{Pines59,Beliaev58};
\begin{eqnarray}
\mu=\frac{\partial E(n_0,\mu)}{\partial n_0},
E(n_0,\mu)=\sum_{M=2}^{\infty}
g_M(n_0,\mu) \frac{n_0^M}{M!},
\label{mu}
\end{eqnarray}
where $g_M(M=2,3,...)$ are the irreducible $M$-body potentials that we will focus on below.
The density of condensed atoms $n_0$ is further constrained by the total number density $n$ as
\begin{eqnarray}
n= n_0- \frac{\partial E(n_0,\mu)}{\partial \mu},
\label{QD}
\end{eqnarray}
In the dilute limit,
the Hartree-Fock energy density is given by Eq.~(\ref{mu}), with
$g_2=4\pi a/m$ and the rest of the potentials $g_{M}, M=3,4...$ set to zero.
The one-loop contributions to $g_M$ for $M=3,4,...$ in Figs. 1(c) and 1(d) all scale like $g_2 \sqrt{na^3}$,
and their sum yields the well-known Lee-Huang-Yang (LHY) correction to the energy density~\cite{Lee57}.
When evaluated in the usual dilute-gas expansion,
$g_{2}$ as well as one-loop contributions formally diverge as $a$ becomes infinite.
Below we regroup these contributions into effective potentials $g_{2,3...}$ at a finite density $n_0$ via resummation of a set of diagrams in the perturbation theory.
The approximation produces a convergent result for $\mu$.

Before proceeding further, we make the following general remark.
In the standard diagrammatic approach~\cite{Beliaev58,Pines59},
the chemical potentials can have contributions from
diagrams with $L$ internal lines, $S$ interaction vertices, and $X$ incoming or outgoing zero momentum lines,
and $X=2S-L$.
For the normal self-energy ($\Sigma_{11}$) and the anomalous counterpart ($\Sigma_{02}$) introduced by Beliaev, by classifying the diagrams
Hugenholtz and Pines had shown that,
in general, the following identity holds~\cite{Pines59}
in the limit of zero energy and momentum:
$\mu=\Sigma_{11}-\Sigma_{02}$. Following a very similar calculation, we further find that
\begin{eqnarray}
\Sigma_{11}(n_0,\mu) =\mu + n_0 \frac{\partial \mu}{\partial n_0},
\label{SE}
\end{eqnarray}
where $\mu={\partial E(n_0,\mu)}/{\partial n_0}$.
The equality in Eq.~(\ref{SE}) is effectively of a hydrodynamic origin.
Following Eq.~(\ref{SE}),
the speed of Bogoliubov phonons~\cite{Bogoliubov47} $v_s$ can be directly related to an {\it effective compressibility}
$\partial n_0/\partial \mu$
via $m v^2_s={\Sigma_{11}-\mu}={n_0}\frac{\partial \mu}{\partial n_0}$,
where the first equality is due to the Hugenholtz-Pines theorem on the phonon spectrum~\cite{HDC}.
Note that hydrodynamic considerations
had also been employed previously by Haldane to construct the
Luttinger-liquid formulation for one-dimensional (1D) Bose fluids~\cite{Haldane81}.
When $na^3$ is small,
Eq.~(\ref{SE}) leads to the well-known result, $\Sigma_{11}=2\mu$.

The self-consistent approach outlined below is mainly suggested by an observation that a subclass of one-loop diagrams [shown in Fig. 1(c)]
yields almost all contributions in the LHY correction (see below and Appendixes A and B).
Resummation of these and their $N$-loop counterparts can be conveniently carried out by introducing the {\it renormalized} or effective
potentials $g_{2,3}$ as shown in Figs. 1(a) and 1(b),
where all internal lines represent,
instead of the noninteracting Green's function
$G_0^{-1}(\epsilon,{\bf k})=\epsilon-\epsilon_{\bf k}+\mu+i\delta$,
the interacting Hartree-Fock Green's function,
$G^{-1}(\epsilon,{\bf k})=\epsilon -\epsilon_{\bf k} -\Sigma_{11}+\mu+i\delta$.
This approximation
captures the main contributions to the chemical potential in the dilute limit
because the renormalization of two-body interactions is mainly
due to virtual states with energies higher than $\mu$
where the Hartree-Fock treatment turns out to be a good approximation.
The self-consistent equation for $\mu$ can be derived by
estimating $g_{2,3,...}(n_0,\mu)$ diagrammatically (see examples in Fig. 1).
When neglecting $g_{3,4,...}$ potentials in Eq.~(\ref{mu}), one obtains
\begin{eqnarray}
\mu &=& n_0 g_2(n_0,\mu)+\frac{n_0^2}{4}g_2^2(n_0,\mu)
\int\frac{d^3{\bf k}}{(2\pi)^3}
\frac{\partial \Sigma_{11}/\partial n_0}{(\epsilon_{\bf k}+\Sigma_{11} -\mu)^2},
\nonumber\\
n &=& n_0+ \frac{n_0^2}{4}g_2^2(n_0,\mu)
\int\frac{d^3{\bf k}}{(2\pi)^3}
\frac{1-\partial \Sigma_{11}/\partial \mu}{(\epsilon_{\bf k}+\Sigma_{11} -\mu)^2},
\nonumber\\
\frac{1}{g_2}&=&\frac{m}{4\pi a}+ \frac{1}{2}
\int \frac{d{\bf k}}{(2\pi)^3}
(\frac{1}{\epsilon_k+\Sigma_{11}
-\mu}-\frac{1}{\epsilon_k}).
\label{SC}
\end{eqnarray}
Equations (\ref{SE}) and (\ref{SC}) can be solved self-consistently.

\begin{figure}
\includegraphics[width=\columnwidth]{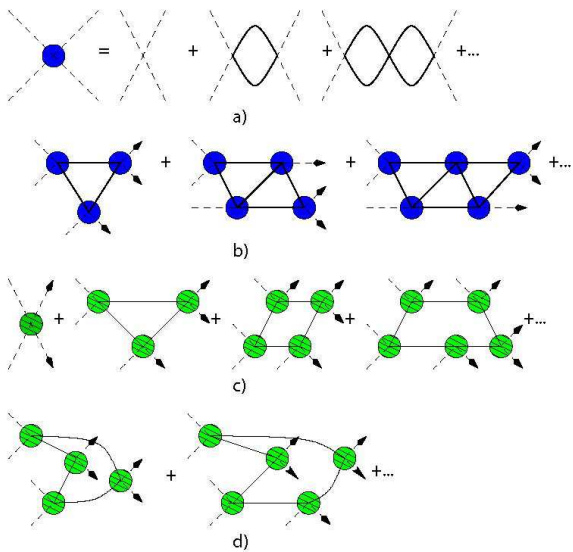}
\caption{ (Color online)
Diagrams showing contributions to the total energy $E(n_0,\mu)$.
The dashed lines are for $k=0$ condensed atoms, thick solid internal lines in (a) and (b) are for interacting
Green's functions $G^{-1}(\epsilon,{\bf k})=\epsilon-\epsilon_{\bf k}-\Sigma_{11}+\mu+i\delta$,
and thin solid lines in (c) and (d) are for noninteracting Green's function
$G_0^{-1}(\epsilon,{\bf k})=\epsilon-\epsilon_{\bf k}+\mu+i\delta$.
(a) The blue circle is for $g_2(n_0,\mu)$;
vertices here represent the bare interaction $U_0$ in Eq.~(1).
(b) ($N=1,2,...$)-loop diagrams that lead to the integral equation for $G_3(-3\eta, p)$ in Eq.~(7).
Note that the usual tree-level diagram violates the momentum conservation and does not exist;
the one-loop diagram has already been included in $g_2(n_0,\mu)$ and therefore needs to be subtracted when calculating
$g_{3}(n_0,\mu)$.
Arrowed dashed lines here as well as in (c) and (d)
stand for outgoing condensed atoms, and the remaining dashed lines stand for incoming ones.
(c) and (d) The tree level and examples of one-loop diagrams that yield
the usual Lee-Huang-Yang corrections in the limit of small $na^3$.
The self-consistent approach contains contributions from (c)-type diagrams but not (d)-type ones
(see further discussion in the text). Patterned green circles also represent the sum of diagrams in (a), but
with thin internal lines, or the noninteracting Green's function $G_0$ lines. All vertices are time ordered from left to right. }
\label{fig0}
\end{figure}

\begin{figure}
\includegraphics[width=\columnwidth]{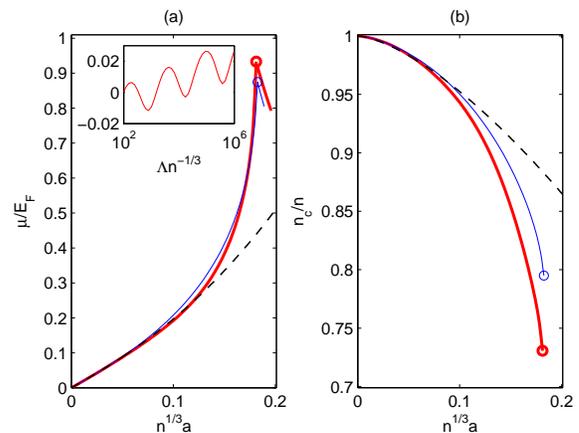}
\caption{(Color online) (a) Chemical potential $\mu$ in units of the Fermi energy $\epsilon_F$ and (b) condensation fraction as a function of
$n^{1/3}a$. Beyond a critical value of $0.18$ (shown as circles),
the solutions become complex, and only the real part of $\mu$ is plotted;
the imaginary part of $\mu$ scales like $\epsilon_F(a/a_{cr}-1)^{1/2}$ near $a_{cr}$.
(However, the sharp transition would be smeared out if the small imaginary part of $G_3$ is included.)
Dashed lines are the result of the Lee-Huang-Yang theory,
thin solid blue lines are the solution without three-body effects (i.e. $g_3=0$).
Thick solid red lines are the solution with $g_3$ included; the momentum cutoff is $\Lambda=100 n^{1/3}$.
The inset is the relative weight of three-body effects in the chemical potential
as a function of $\Lambda n^{-1/3}$ at the critical point.
}
\label{fig1}
\end{figure}

We first benchmark our results with the LHY correction or Beliaev's
results for $\mu$ by solving the equations in the limit of small $na^3$. We find
$\mu=\frac{4\pi}{m} n_0 a (1+ 3\sqrt{2\pi}\sqrt{n_0 a^3}+...)$, and
the number equation yields an estimate
$n_0/n=(1-\frac{\sqrt{2\pi}}{2}\sqrt{na^3}+...)$.
The second terms in the parentheses
are of the same nature as the LHY correction.
Comparing to  Beliaev's  perturbative result for chemical potential,
$\mu=\frac{4 \pi}{m}n_0 a (1+ \frac{40}{3\sqrt{\pi}} \sqrt{n_0 a^3}+...)$~\cite{Beliaev58}, and for the condensation fraction
$n_0/n=1- \frac{8}{3\sqrt{\pi}}\sqrt{na^3}+...$,
one finds that the self-consistent solution reproduces
$99.96\%(=9\pi\sqrt{2}/40)$
of the Beliaev's correction for the chemical potential,
and $83.30\% (=3\pi\sqrt{2}/16)$ of the depletion fraction in the dilute limit.
Technically,
one can further examine $g_2(n_0,\mu)$ by expanding it in terms of $a$ and $\Sigma_{11}$ and
then compare with
the usual diagrams in the dilute gas theory~\cite{Beliaev58}.
One indeed finds that $g_2(n_0,\mu)$ in
Eq.~(\ref{SC}) effectively includes
{\it all} one-loop diagrams with
$X=3,4,5,...$ incoming or outgoing zero-momentum lines
that involve a {\it single pair} of virtually excited atoms [between any two consecutive scattering vertices; Fig. 1(c)]. The one-loop diagrams with
$X=4,5,...$ incoming or outgoing zero-momentum lines that involve multiple pairs of virtual
atoms [Fig. 1(d)] have been left out, but they only count for less than $0.04\%$ of Beliaev's result~\cite{G4}.

Following the same line of thought, one can also verify that
$g_{2}(n_0,\mu)$ further contains ($N=2,3,4,..$)-loop contributions
that only involve {\it one pair} of virtual atoms;
each two adjacent loops only share one interaction vertex and are reducible.
$g_3(n_0,\mu)$ included below, on the other hand, includes ($N=2,3,4,..$)-loop contributions with
$S=4,5...$ interaction vertices that only involve three virtual atoms;
two adjacent loops share one internal line instead of a single vertex [see Fig. 1(b)]
and are irreducible, {\it i.e.}, cannot be expressed as a simple product of individual loops.
Effectively, we take into account all the virtual processes involving either two or three dressed excited atoms in the calculation of
the chemical potential $\mu$ by including the effective $g_{2,3}$ (defined in Fig. 1) in Eq.~(\ref{mu}). The processes involving
four or more excited atoms only appear in $g_{4,5...}$ and are not included here; at the one-loop level following
the above calculations, the corresponding contributions from the processes involving multiple pairs of virtual atoms are indeed negligible.

A solution to Eq.~(\ref{SC}) is shown in Fig. 2.
An interesting feature of Eq.~(\ref{SC}) is that it no longer has a real
solution once $n^{1/3}a$ exceeds the critical value of $0.18$, implying an onset instability; this is not anticipated in the dilute-gas theory~\cite{Lee57}.
This can also be illustrated by considering the two-body effective coupling constant
$G_2(\Lambda_0)$ as a function of $\Lambda_0$~\cite{Cui10},
a characteristic momentum that defines a low-energy subspace,
\begin{eqnarray}
G_2(\Lambda_0)=\frac{4\pi}{m}\frac{1}{\frac{1}{a}-\frac{2}{\pi}\Lambda_0}.
\end{eqnarray}
For Bose gases, it is appropriate to identify the relevant $\Lambda_0$ as $\sqrt{2m\mu}=\Lambda_{\mu}$.
For positive scattering lengths, $a$ not
only defines the strength of interaction
in the small $\Lambda_\mu$ or dilute limit but also
sets a scale for $\Lambda_\mu$, above which the effective interaction becomes negative, {\it i.e.},
$G_2 (\Lambda_\mu)< 0$ if $\Lambda_\mu > \pi/(2a)$.
So as $a$ approaches infinity, condensed atoms with a chemical potential $\mu$ typically see each other as attractive
rather than repulsive, resulting in molecules~\cite{Upper}.
Thus, beyond the critical point the upper branch atomic gases become strongly coupled to the molecules with a strength proportional to the imaginary part of $\mu$.
Consequently, we anticipate that $\mu$ decreases quickly beyond the critical scattering length due to the formation of molecules, leading to a maximum in $\mu$ in the vicinity of the critical
point~\cite{data}.

A renormalization group approach based on atom-molecule fields
was also applied in a previous study to understand
Bose gases near resonance~\cite{Lee10,FTA}.
Our results differ from theirs
in two aspects. First, in our approach, an onset instability
sets in near resonance even when the scattering length is positive, a key feature that is absent in that previous study.
Second, when extrapolated to the limit of small $na^3$,
the results in Ref.~\cite{Lee10} imply a
correction of the order of $\sqrt{na^3}$ to
the usual Hartree-Fock chemical potential but
with a negative sign, opposite to the sign of LHY corrections.
In a recent study~\cite{Diederix11},
a self-consistent mean-field equation was employed, leading to a similar conclusion as the approach in Ref.~\cite{Lee10}; the approach
does not yield the correct sign of the LHY corrections.
And so the onset instability pointed out in this paper, which is surprising from the point of view of dilute-gas theory, is also absent there.

The chemical potential near the critical point can be estimated
using Eq.~(\ref{SC}) and is close to $0.9\epsilon_F$, where $\epsilon_F=(6\pi^2)^{2/3}n^{2/3}/2m$ is the Fermi energy defined for a gas of density $n$.
This is consistent with the picture of nearly fermionized Bose gases suggested by the previous calculations and experiments~\cite{Cowell02,Song09,Lee10,Diederix11,Navon11}.

\begin{figure}
\includegraphics[width=\columnwidth]{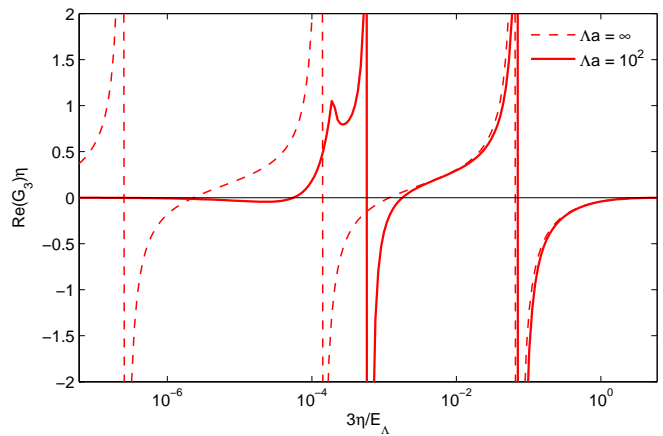}
\caption{
(Color online) ${\rm Re} G_3(-3\eta,0)\eta$ as a function of $\eta=\Sigma_{11}(n_0)-\mu$.
$E_\Lambda=\Lambda^2/2m$ and
$\Lambda$ is the momentum cut-off.
The imaginary part of $G_3$ (not shown) is zero once $3\eta > 1/(ma^2)$.
}
\label{fig3}
\end{figure}

We now turn to the effect of $g_3(n_0,\mu)$ on the chemical potential by including it in Eq.~(\ref{mu}).
We estimate $g_3$ by summing up all $N$-loop diagrams with
$X=3$ incoming or outgoing zero momentum lines,
which are represented in Fig. 1. All diagrams have three incoming or outgoing zero momentum lines but with $N=2,3,..$ loops.
The effect of three-body forces
due to Efimov states~\cite{Efimov70}
was previously studied
in the dilute limit~\cite{BHM02}.
The deviation of the energy density from the usual universal
structures (i.e., only depends on $na^3$ ) was obtained by studying
the Efimov forces in the zero-density limit.
The contribution obtained there scales like $a^4$, apart from a log-periodic modulation~\cite{BHK99}, and again formally diverges as other terms
when approaching a resonance. There
was also an interesting proposal of a liquid-droplet phase at
negative scattering lengths but in the vicinity of a trimer-atom threshold~\cite{Bulgac02}.

It is necessary to regularize the
usual $a^4$ behavior at resonance in the three-body forces by further taking into account
the interacting Green's function when calculating the $N$-Loop six-point correlators.
Including the self-energy in the calculation,
we remove the $a^4$ dependence that usually appears in the Bedaque-Hammer-Van Kolck theory for the three-body forces~\cite{BHK99};
when setting $\mu,\Sigma_{11}$ to zero, the equation collapses into
the corresponding equation for three Bose atoms in vacuum, which was previously employed to obtain the $\beta$ function for the renormalization flow in an atom-dimer field-theory model.
The sum of loop diagrams in Fig. 1(b), $G_3(-3\eta,p)$, satisfies a simple integral equation ($m$ set to be unity; see Appendix C):
\begin{eqnarray}
& &G_3(-3\eta,p)=
\frac{2}{\pi} \int dq
K(-3\eta;p,q)  \nonumber \\
& & \times \frac{q^2}{\sqrt{\frac{3q^2}{4}+3\eta}-\frac{1}{a}}
[ G_3(-3\eta,q)-\frac{1}{q^2+3\eta} ],
 \nonumber\\
& & K(-3\eta;p,q)=\frac{1}{pq}\ln \frac{p^2+q^2+pq+3\eta}{p^2+q^2-pq+3\eta},
\label{3body}
\end{eqnarray}
where we have introduced $\eta=\Sigma_{11}(n_0)-\mu$.
$G_3(-3\eta,0)$ is plotted numerically in Fig. 3. Three-body potential $g_3(n_0,\mu)$ is related to $G_3(-3\eta,0)$ via
$g_3(n_0,\mu)= 6 g^2_2 {\rm Re} \tilde{G}_3(-3\eta,0)$ where $\tilde{G}_3$ is obtained by further subtracting from $G_3$
the one-loop diagram in Fig. 1(b)
because its contribution has already been included in $g_2(n_0,\mu)$.
The structure of $G_3(-3\eta,0)$ is particularly simple
at $a=+\infty$, as shown in Fig. 3: It has a desired log-periodic behavior reflecting the underlying
Efimov states~\cite{Efimov70}.
When $3\eta$ is close to an Efimov eigenvalue $B_n=B_0 \exp(-2\pi n/s_0)$ [$n=1,2,3....$,$\exp(2\pi/s_0)=515$]
that corresponds to a divergence
point in Fig. 3, the three-body forces are the most significant.
When $3\eta$ is in the close vicinity of zeros in Fig. 3,
the three-body forces are the negligible
and Bose gases near resonance are dictated by the $g_2$ potential.

When including the real part of $g_3(n_0,\mu)$ in the calculation of $E(n_0,\mu)$, we further get an estimate of
three-body contributions to the energy density and chemical potential $\mu$.
The contribution is nonuniversal and depends on the
momentum cutoff in the problem.
For typical cold Bose gases, it is reasonable to assume the momentum cutoff $\Lambda$ in the integral equation Eq.~(\ref{3body})
to be $100 n^{1/3}$ or even larger. Quantitative effects on the chemical potential are presented in Fig. 2.

Note that $G_3(-3\eta,0)$ also has an imaginary part even at small scattering lengths; this corresponds to the well-known
contribution of three-body recombination. The onset instability discussed here will be further rounded off if the imaginary part of $G_3$ is included.
However, for the range of parameters we studied, both the real and imaginary parts of $G_3$ appear to be numerically small (see also Fig. 2);
the energetics and instabilities near $a_{cr}$ are found to be mainly determined by the renormalized two-body interaction $g_2(n_0,\mu)$.

\section{Conclusions}
In conclusion, we have investigated the energetics of Bose gases near resonance beyond the Lee-Huang-Yang dilute limit via a simple resummation scheme.
We have also pointed out an onset instability and estimated three-body Efimov effects that had been left out
in recent theoretical studies of Bose gases near resonance~\cite{Cowell02,Song09,Lee10,Diederix11}.
Within our approach, we find that the three-body forces contribute around a few percent to the chemical potential and that
the Bose gases are nearly fermionized before an onset instability sets in near resonance.

\section*{acknowledgement}
This work is in part supported by Canadian Institute for Advanced Research,
Izzak Walton Killam Foundation, NSERC (Canada), and the Austrian Science Fund FWF FOCUS.
One of the authors (F.Z.) also would like to thank the Institute for Nuclear Physics,
University of Washington, for its hospitality during a cold-atom workshop
in Spring, 2011.
This work was prepared at the Aspen center for physics during
the 2011 cold-atom workshop.
We would like to thank
Aurel Bugalc,
Eric Braaten, Randy Hulet, Gordon Semenoff, Dam T. Son, Shina Tan, Lan Yin and Wilhelm Zwerger for helpful discussions.

\begin{widetext}
\appendix
\section{Solving Self-consistent Equation (5) in The Dilute Limit}

We apply Eq.~(5) to calculate the leading-order correction beyond the
mean-field theory. We notice that the equations for $g_{2}$ and $\mu$
are arranged in a way that the next-order correction can be obtained
by applying the results from the lowest-order approximation to the
right-hand side. In the lowest-order approximation, we find $\Sigma_{11}=8\pi n_{0}a$
and $\mu=4\pi n_{0}a$; this leads to a correction to $g_{2}$ as
\begin{eqnarray}
g_{2} & = & 4\pi a+\frac{\left(4\pi a\right)^{2}}{2}\int\frac{d^{3}k}{\left(2\pi\right)^{3}}\left(\frac{1}{\epsilon_{k}}-\frac{1}{\epsilon_{k}+\mu}\right)=4\pi a\left(1+\sqrt{8\pi n_{0}a^{3}}\right).\label{eq:g2-exp-sc}
\end{eqnarray}
Similarly, from the relation $\frac{\partial\Sigma_{11}}{\partial n_{0}}=8\pi a$
and $\frac{\partial\Sigma_{11}}{\partial\mu}=0$, we can get the correction
for the chemical potential $\mu$ as,
\begin{eqnarray}
\mu & = & 4\pi an_{0}+\frac{\left(4\pi a\right)^{3}n_{0}^{2}}{2}\int\frac{d^{3}k}{\left(2\pi\right)^{3}}\frac{1}{\left(\epsilon_{k}+\mu\right)^{2}}=4\pi an_{0}\left(1+3\sqrt{2\pi n_{0}a^{3}}\right)\label{eq:mu-leading-sc}
\end{eqnarray}
 and the depletion fraction
\begin{equation}
\frac{n_{p}}{n_{}}=\frac{n_{0}}{4}g_{2}^{2}\int\frac{d^{3}k}{\left(2\pi\right)^{3}}\frac{1}{\left(\epsilon_{k}+\mu\right)^{2}}=\sqrt{\frac{\pi}{2}n_{0}a^{3}}.
\end{equation}

For a comparison we list the results from the dilute-gas theory,
\begin{eqnarray}
\mu_{Beliaev} & = & 4\pi n_{0}a\left[1+\frac{40}{3}\sqrt{\frac{1}{\pi}n_{0}a^{3}}\right],\\
\left(\frac{n_{p}}{n_{}}\right)_{Beliaev} & = & \frac{8}{3}\sqrt{\frac{1}{\pi}n_{0}a^{3}}.
\end{eqnarray}
 Our self-consistent approach produces $\frac{9\sqrt{2}\pi}{40}(=99.96\%)$
of Beliaev's result for the chemical potential, and $\frac{3\sqrt{2}\pi}{16}(=83\%)$
for the depletion fraction.

\section{A Comparison Between the Self-Consistent Approach and the Dilute-
Gas Theory}

In the following, we show explicitly that our self-consistent equation
corresponds to a subgroup of diagrams [in Fig. 1(c)] in the usual dilute
gas theory. The two-body $T$-matrix used in the dilute-gas theory
[represented by the green circles in Figs. 1(c) and 1(d)] are obtained using the
non-interacting Green's function $G^{{-1}}(\epsilon,k)=\epsilon-\epsilon_{k}+\mu+i0^{+}$;
in the dilute limit, we can expand the $T$-matrix as
\begin{equation}
t(\omega,Q)=4\pi a\left[1+4\pi a\int\frac{d^{3}k}{\left(2\pi\right)^{3}}\left(\frac{1}{\omega-\frac{Q^{2}}{4}-k^{2}+2\mu+i0^{+}}+\frac{1}{k^{2}}\right)+\cdots\right],
\end{equation}
where $\omega$ and $Q$ are the total energy and momentum of the
incoming atoms. The contribution from the first two diagrams in Fig. 1(c)
are
\begin{eqnarray}
E_{(c1)} & \simeq & \frac{t(0,0)n_{0}^{2}}{2}\simeq2\pi an_{0}^{2}\left[1+4\pi a\int\frac{d^{3}k}{\left(2\pi\right)^{3}}\left(\frac{1}{-k^{2}+2\mu+i0^{+}}+\frac{1}{k^{2}}\right)\right]\nonumber \\
E_{(c2)} & \simeq & 2\frac{t^{2}(0,0)n_{0}^{2}}{2}\int\frac{d^{3}k}{\left(2\pi\right)^{3}}\left(\frac{1}{-k^{2}+2\mu+i0^{+}}\right)^{2}2n_{0}t(\mu-\epsilon_{k},0)\\
 & \simeq & 2\pi an_{0}^{2}\left[(4\pi a)\left(16\pi n_{0}a\right)\int\frac{d^{3}k}{\left(2\pi\right)^{3}}\left(\frac{1}{-k^{2}+2\mu+i0^{+}}\right)^{2}\right].\label{eq:c2}
\end{eqnarray}
For the leading-order correction beyond the mean-field theory, it
suffices to set $t(\mu-\epsilon_{k},0)\simeq4\pi a$ in Eq.~(\ref{eq:c2})
and in higher-order diagrams. Similarly, we can get the contributions
from the higher-order diagrams in this series, and the sum is
\begin{eqnarray}
E_{(c)} & \simeq & 2\pi an_{0}^{2}\left[1+4\pi a\int\frac{d^{3}k}{\left(2\pi\right)^{3}}\left(\frac{1}{-k^{2}+2\mu+i0^{+}}+\frac{1}{k^{2}}\right)\right]\nonumber \\
 & + & 2\pi an_{0}^{2}(4\pi a)\sum_{m=1}^{\infty}\left(16\pi an_{0}\right)^{m}\int\frac{d^{3}k}{\left(2\pi\right)^{3}}\left(\frac{1}{-k^{2}+2\mu+i0^{+}}\right)^{m+1}\label{eq:Ec-series}\\
 & \simeq & 2\pi an_{0}^{2}\left[1+4\pi a\int\frac{d^{3}k}{\left(2\pi\right)^{3}}\left(\frac{1}{-k^{2}+2\mu-16\pi n_{0}a}+\frac{1}{k^{2}}\right)\right].\label{eq:Ec-all}
\end{eqnarray}
We see that the energy given by the diagrams in Fig. 1(c) is \emph{exactly}
the same as the one used in our self-consistent equation, e.g., $g_{2}n_{0}^{2}/2$,
where $g_{2}$ should be expanded as Eq.~(\ref{eq:g2-exp-sc}) in the
dilute limit.

Next, we can sum up the rest of the one-loop diagrams that are not included
in the self-consistent equations; they represent the lowest-order
contributions to four- and six-body forces and so on. In the dilute limit,
these diagrams [as shown in Fig. 1(d)] can be summed as
\begin{eqnarray}
E_{(d)} & = & -\left(4\pi an_{0}\right)\int\frac{d^{3}k}{\left(2\pi\right)^{3}}\sum_{m=2}^{\infty}\frac{1}{2}\frac{\left(2m-2\right)!}{m!\left(m-1\right)!}\left(\frac{4\pi an_{0}}{2\epsilon_{k}-2\mu+16n_{0}\pi a}\right)^{2m-1}.
\end{eqnarray}
 Indeed, we can recover Beliaev's result by summing up one-loop diagrams
in Figs. 1(c) and 1(d) as
\begin{eqnarray}
\frac{\partial}{\partial n_{0}}\left(E_{(c)}+E_{(d)}\right) & = & 4\pi n_{0}a+4\pi a\int\frac{d^{3}k}{\left(2\pi\right)^{3}}\left[\frac{\left(\epsilon_{k}-\mu+6\pi n_{0}a\right)}{\sqrt{\left(\epsilon_{k}-\mu+8\pi an_{0}\right)^{2}-\left(4\pi an_{0}\right)^{2}}}-1+\frac{4\pi n_{0}a}{k^{2}}\right]\\
 & = & 4\pi n_{0}a\left[1+\frac{40}{3}\sqrt{\frac{1}{\pi}n_{0}a^{3}}\right]=\mu_{Beliaev}
\end{eqnarray}

\section{Including Three-body Forces In the Self-Consistent Equations}

We now calculate the amplitude of three-body scatterings corresponding to the processes described in Fig. 1(b).
First, we consider a general case where three incoming momenta, instead of being zero, are
${\bf k}_1={\bf p}/2-{\bf q}$, ${\bf k}_2={\bf p}/2+{\bf q}$, and
${\bf k}_3=-{\bf p}$, and the outgoing ones are
${\bf k}'_1={\bf p}'/2-{\bf q}'$, ${\bf k}'_2={\bf p}'/2+{\bf q}'$, and
${\bf k}'_3=-{\bf p}'$.
The scattering amplitude between theses states is then given by $A(E-3\eta;{\bf p}, {\bf p}')$,
which represents the sum of diagrams identical to Fig. 1(b) except that the external lines carry finite momenta.

For the estimate of three-body contributions of $g_3$, we first treat
the sum of diagrams in Fig. 1(b) as the limit of $A(E-3\eta;{\bf p},{\bf p'})$ when ${\bf p}$ and ${\bf p}'$ approach zero and
the total frequency $E$ is set to zero.
It is therefore more convenient to work with the reduced amplitude
$G_3(E-3\eta;{\bf p})=A(E-3\eta;{\bf p},0)$, where ${\bf p}'$ is already taken to be zero.
$G_3(E-3\eta;{\bf p})$ itself obeys a simple integral equation, as can be seen by listing the terms in the summation explicitly.
Indeed, when $E$ is further set to zero, we find that the diagrams in Fig. 1(b) yield
\begin{eqnarray}
G_{3}(-3\eta,p) & = & \frac{2}{\pi}\int dq\frac{K(-3\eta;p,q)q^{2}}{\sqrt{\frac{3}{4}q^{2}+3\eta}-\frac{1}{a}}\frac{-1}{q^{2}+3\eta}\\
 & + & \left(\frac{2}{\pi}\right)^{2}\int dqdq'\frac{K(-3\eta;p,q)q^{2}}{\sqrt{\frac{3}{4}q^{2}+3\eta}-\frac{1}{a}}\frac{K(-3\eta;q,q')q'^{2}}{\sqrt{\frac{3}{4}q'^{2}+3\eta}-\frac{1}{a}}\frac{-1}{q'^{2}+3\eta}+\cdots,
\label{sum1}
\end{eqnarray}
where $K(-3\eta;p,q)$ is the kernel defined in the main text. The
sum of the above infinite series leads to the following integral equation
of $G_{3}$ :
\begin{equation}
G_{3}(-3\eta,p)=\frac{2}{\pi}\int dq\frac{K(-3\eta;p,q)q^{2}}{\sqrt{\frac{3}{4}q^{2}+3\eta}-\frac{1}{a}}\frac{-1}{q^{2}+3\eta}+\frac{2}{\pi}\int dq\frac{K(-3\eta;p,q)q^{2}}{\sqrt{\frac{3}{4}q^{2}+3\eta}-\frac{1}{a}}G_{3}(-3\eta,q).
\end{equation}
Note that $G_{3}(-3\eta,0)$ defined above includes a diagram [the leftmost
one in Fig. 1(b)] that has already been included in $g_{2}$. To avoid
overcounting, we subtract the first diagram in Fig. 1(b) from $G_{3}$
as
\begin{eqnarray}
g_{3} & = & 6 g_{2}^{2}{\rm Re}\left[G_{3}(-3\eta,0)-\frac{2}{\pi}\int dq\frac{K(-3\eta;0,q)q^{2}}{\sqrt{\frac{3}{4}q^{2}+3\eta}-\frac{1}{a}}\frac{-1}{q^{2}+3\eta}\right].
\end{eqnarray}

Alternatively, one can also carry out
a direct summation of the diagrams in Fig. 1(b). It leads to a result that numerically differs very little from the
estimation obtained above via an asymptotic extrapolation. For instance, a direct evaluation of those diagrams yields
\begin{eqnarray}
G_{3}(-3\eta,0) & = & \frac{2}{\pi}\int dq\frac{K(-2\eta;0,q)q^{2}}{\sqrt{\frac{3}{4}q^{2}+3\eta}-\frac{1}{a}}\frac{-1}{q^{2}+2\eta}\\
& + & \left(\frac{2}{\pi}\right)^{2}\int dqdq'\frac{K(-2\eta;0,q)q^{2}}{\sqrt{\frac{3}{4}q^{2}+3\eta}-\frac{1}{a}}\frac{K(-3\eta;q,q')q'^{2}}{\sqrt{\frac{3}{4}q'^{2}+3\eta}-\frac{1}{a}}\frac{-1}{q'^{2}+2\eta}+\cdots
\label{sum2}
\end{eqnarray}
The only difference between Eqs. (\ref{sum2}) and (\ref{sum1}) is that the frequencies appearing in the first kernel $K(E;0,q)$ in the integrands and
in the last denominators are now $-2\eta$ instead of $-3\eta$.

One can easily verify that the sum can be written in the following compact form:
\begin{equation}
{G}_{3}(-3\eta,0)=\frac{2}{\pi}\int dq\frac{K(-2\eta;0,q)q^{2}}{\sqrt{\frac{3}{4}q^{2}+3\eta}-\frac{1}{a}}
[\frac{-1}{q^{2}+2\eta}+ G^{'}_{3}(-3\eta,q)],
\end{equation}
where $G^{'}_{3}(-3\eta, p)$ is a solution of the following integral equation:
\begin{equation}
G^{'}_{3}(-3\eta,p)=\frac{2}{\pi}\int dq\frac{K(-3\eta;p,q)q^{2}}{\sqrt{\frac{3}{4}q^{2}+3\eta}-\frac{1}{a}}
[\frac{-1}{q^{2}+2\eta}+ G^{'}_{3}(-3\eta,q)]
\end{equation}
Note that $G^{'}_{3}(-3\eta,p)$ defined here describes an off-shell scattering between three incoming atoms with momenta
${\bf p}/2-{\bf q}$, ${\bf p}/2+{\bf q}$, and $-{\bf p}$ and three condensed atoms. And $G_3(-3\eta,0)$ is not equal to $G^{'}_3(-3\eta,0)$, a consequence of the Hartree-Fock approximation
we have employed here. $G_3(-3\eta, 0)$ and $G^{'}_3(-3\eta,0)$ can be obtained numerically.

Finally, after subtracting the leftmost one-loop diagram in Fig. 1(b) we again find the three-body contribution is

\begin{eqnarray}
g_{3} & = & 6 g_{2}^{2}{\rm Re} \left[ {G}_{3}(-3\eta,0)
-\frac{2}{\pi}\int dq\frac{K(-2\eta;0,q)q^{2}}{\sqrt{\frac{3}{4}q^{2}+3\eta}-\frac{1}{a}}\frac{-1}{q^{2}+2\eta}\right].
\end{eqnarray}

Now we can include the three-body forces $\frac{g_{3}n_{0}^{3}}{6}$
in a set of differential self-consistent equations similar to Eq.~(5).
We solve the equation numerically, and the results are shown in Fig. 2,
where in the inset we show the momentum cutoff $\Lambda$ dependence
in the chemical potential. In our numerical program, we further use
the approximation $\frac{\partial\Sigma}{\partial n_{0}}=2g_{2}$,
$\frac{\partial\Sigma}{\partial\mu}=0$, and $\Sigma_{11}=\beta\mu$
($\beta=2$) to simplify the numerical calculations. We have tested
other types of approximation schemes for the self-energy, such as
$\Sigma_{11}=8\pi an_{0}$ or $\Sigma_{11}=2g_{2}n_{0}$. We find
that the chemical potential and the value of the critical point $na_{cr}^{3}$
are insensitive to approximation schemes.

\end{widetext}


\begin{thebibliography}{10}
\bibitem{twobody}
Resummation is possible
for two scattering atoms in a box of size $L$.
The interaction energy should scale like
$\frac{4\pi a}{m L^3}(1+ A \frac{a}{L}+...)$
when $a$ is much less than  $L$ ($A$ is a constant) but generically saturate at a value of the order of
$1/2 m L^2$ when $a$ becomes infinite.
\bibitem{HDC}
If we attribute the energy density to the zero-point energy of Bogoliubov phonons,
for an arbitrary scattering length, one can,
using Eqs. (\ref{QD}) and (\ref{SE}), express the condensation density as
$n_0=n - 1/3\pi^2 (\partial \mu/\partial \ln n_0)^{3/2} m^{3/2}$.
The long-wavelength dynamics thus sets an upper bound
on the value of $\partial\mu/\partial \ln n_0$.
The upper bound can be estimated to be $ 2^{1/3} \epsilon_F$,
where $\epsilon_F=(6\pi^2)^{2/3}n^{2/3}/2m$ is the Fermi energy defined for a gas of density $n$.
\bibitem{data}
Interestingly, a maximum in the Bragg line shift at a finite wave vector was found when the LHY correction is $0.22$ in Ref. \cite{Papp08}.
The maximum in $\mu$ in this paper occurs when the LHY correction is $0.45$.
The implication of the maximum found in this paper on the line shift data will be further studied in the future.
\end{thebibliography}
\end{document}